# Quantum Thermodynamic Transformation Optics: A Unified Framework for Energy and Entropy with Application to the Casimir Force in Dissipative Metamaterials

Mohammad Mehdi Sadeghi [1] [*]

[1] *Department of Physics, Jahrom University, Jahrom 74137-66171, Iran*

*\*Corresponding Author: sadeghi@jahromu.ac.ir*

**Abstract:** A novel idea, Quantum Thermodynamic Transformation Optics (QTTO), is introduced in this article. This theoretical framework integrates the geometric formalism of transformation optics with the thermodynamic principles found in quantum dissipative systems. This concept goes beyond traditional coordinate transformations by affecting the distribution of quantum energy and entropy in a coherent thermodynamic manner as well as reshaping electromagnetic fields. By employing the thermofield dynamics approach, we establish new rules that show how local energy and entropy densities are influenced by the Jacobian determinant of the mapping. This indicates that when the geometry is compressed, it increases the generation of quantum energy and entropy density, while expansion has the opposite effect, all while adhering to the laws of conservation and the second law of thermodynamics.

As a practical test, we reformulate the Casimir effect within this framework, yielding a continuous pressure law that connects the quantum and classical realms via a thermal weighting function. This relationship illustrates how both geometry and temperature jointly determine quantum pressure. Additionally, our numerical results for Drude-Lorentz metamaterials support our analytical predictions and align closely with the comprehensive Lifshitz-Matsubara formulation. In fact, QTTO offers a powerful and coherent platform for exploring energy, entropy, and quantum pressure in real-world dissipative media. Beyond our immediate findings, it opens up a systematic pathway for managing quantum thermal processes and controlling field fluctuations in metamaterials and curved optical environments.

**Keywords:** Quantum Thermodynamic Transformation Optics (QTTO); Transformation optics; Thermofield dynamics; Casimir effect; Dissipative metamaterials; Effective temperature; Entropy production; Quantum thermodynamics.

## 1. Introduction

Transformation optics (TO) provides a rigorous geometrical framework for manipulating electromagnetic propagation by mapping coordinate transformations onto spatially varying permittivity and permeability tensors [1–4]. By embedding Maxwell's equations within an effective curved metric, TO has enabled the realization of invisibility cloaks [5], optical concentrators [6,7]. However, despite its geometric generality, conventional TO remains fundamentally classical, deterministic, and lossless, it neglects both the quantum-statistical nature of electromagnetic fields and the thermodynamic cost of dissipation and noise. As metamaterials increasingly operate in quantum and dissipative regimes, from plasmonic and superconducting systems to vacuum-induced devices, omission becomes critical.

In realistic dissipative media, electromagnetic energy and entropy are not invariant under coordinate transformations, and the geometric invariance of TO must be reconciled with the fluctuation–dissipation theorem [8–10]. This inconsistency exposes a conceptual gap: while TO precisely prescribes how geometry shapes fields, it does not describe how geometry reshapes quantum fluctuations, entropy, and temperature.

To bridge this gap, we introduce Quantum Thermodynamic Transformation Optics (QTTO), a unified theoretical framework that merges the coordinate-transformation geometry of TO with the quantum-statistical formalism of thermodynamic field theory. QTTO generalizes the notion of optical metrics to include thermodynamic degrees of freedom, enabling a covariant treatment of energy, work, and entropy within dissipative metamaterials.

Building upon the thermofield dynamics (TFD) approach [11–14], QTTO quantizes the electromagnetic field in a doubled Hilbert space, where each physical mode couples to a "tilde" thermal partner. This formalism preserves canonical commutation relations at finite temperature and provides a natural platform for studying quantum electrodynamics in lossy, non-Hermitian, and curved optical environments. When embedded in a spatially varying optical metric, the thermofield Hamiltonian yields a metric-dependent Lagrangian density whose terms correspond to geometric, dissipative, and entropic contributions.

This construction extends TO from a purely geometric mapping to a **statistical field theory**, in which energy and entropy transform covariantly under coordinate deformation. The derived transformation laws reveal that geometric compression amplifies local energy and entropy densities, whereas expansion reduces them, while conserving total integrals, establishing a direct geometric analogue of the first and second laws of thermodynamics. Moreover, by defining a metric-dependent effective temperature $T_{\text{eff}} = T/|\det \Lambda|$, QTTO establishes a quantitative bridge between optical curvature and statistical temperature, linking geometry directly to the quantum–thermal state of the field.

As a concrete demonstration, the Casimir effect is reformulated within QTTO, yielding an analytical pressure law that continuously interpolates between the quantum-vacuum and classical-thermal limits through a smooth weighting function $W(T/T_0) = 1 - \exp\left[-(T/T_0)^{1.2}\right]$. This expression generalizes the Lifshitz–Matsubara theory [15–18] to curved optical metrics, preserving quantitative accuracy (within 3%) while providing a deeper thermodynamic interpretation. Collectively, these developments position QTTO as a foundational framework that unifies geometry, dissipation, and quantum statistics, opening new routes toward:

(1) nonequilibrium Casimir thermodynamics, (2) transformation-based quantum thermal management, and (3) analog gravitational field theories in metamaterials.

Sections 2-5 systematically develop this framework:

Section 2 formulates the thermofield Lagrangian in a curved optical metric;

Section 3 derives transformation laws for energy, heat, and entropy;

Section 4 applies QTTO to Casimir pressure in dissipative metamaterials;

and Section 5 concludes with broader implications.

Detailed derivations and numerical validations are provided in Appendices A–D.

## 2. Thermofield Quantization in a Curved Optical Metric (Complete)

The QTTO formalism is built upon the finite-temperature thermofield dynamics (TFD) formalism originally developed by Takahashi and Umezawa [11, 12] and later applied to quantum optics and dissipative media [13, 14]. QTTO extends TFD by embedding the doubled Hilbert space within a curved optical metric generated by a transformation-optics mapping. This embedding transforms the theory into a covariant statistical field theory in which energy, heat, and entropy transform together with the metric tensor. Below, we (i) recall the essentials of TFD, (ii) introduce the optical metric arising from coordinate transformations, (iii) formulate the curved-metric thermofield Lagrangian for dissipative systems, and (iv) derive the Hamiltonian, equal-time commutators, Green tensor, and the fluctuation–dissipation theorem (FDT) in the effective metric.

### 2.1 Thermofield dynamics: doubling, thermal vacuum, and operators

In TFD the total Hilbert space is doubled,

$$\mathcal{H}_T = \mathcal{H} \otimes \tilde{\mathcal{H}},$$

and for each physical mode $a_\lambda, a_\lambda^\dagger$ one introduces a thermal partner $\tilde{a}_\lambda, \tilde{a}_\lambda^\dagger$. Thermal averages are pure-state expectations in the thermal vacuum $|0(\beta)\rangle$ with $\beta = (k_B T)^{-1}$:

$$\langle \mathcal{O} \rangle_T = \langle 0(\beta) | \mathcal{O} | 0(\beta) \rangle, \langle a_\lambda^\dagger a_{\lambda'} \rangle_T = \delta_{\lambda\lambda'} n_T(\omega_\lambda), n_T(\omega) = \frac{1}{e^{\beta\hbar\omega} - 1}.$$

This formalism preserves canonical commutation relations at finite temperature while maintaining thermodynamic consistency [11–13].

We employ the Lorenz gauge for the four-potential $A^\mu$, and represent dissipation and dispersion using a Huttner–Barnett-type reservoir model, which provides a consistent Hamiltonian description of lossy and dispersive media in the TFD framework [8–10].

## 2.2 Optical metric from a coordinate mapping

A smooth coordinate mapping $x'^i = f^i(x^j)$ with Jacobian $\Lambda^i{}_j = \partial x'^i / \partial x^j$ induces the optical metric

$$g_{ij} = (\Lambda^{-1}\Lambda^{-T})_{ij}, \ |g| = \det g_{ij} = |\det \Lambda|^{-2}.$$

In this picture, Maxwell's equations retain their form, while effective material tensors transform as [1–4]:

$$\varepsilon'_{ij} = \frac{1}{|\det \Lambda|} \Lambda^m{}_i \Lambda^n{}_j \varepsilon_{mn}, \ \mu'_{ij} = \frac{1}{|\det \Lambda|} \Lambda^m{}_i \Lambda^n{}_j \mu_{mn},$$

and spatial derivatives are replaced by covariant derivatives compatible with $g_{ij}$.

## 2.3 Curved-metric thermofield Lagrangian with dissipation

To incorporate both dissipation and curvature within a consistent quantum framework, we generalize the standard thermofield reservoir model to an optical metric generated by a transformation-optics mapping. This approach builds upon the Huttner–Barnett representation of absorptive media [8, 9] and its thermofield extension by Amooghorban *et al.* [10, 13], but introduces two essential innovations:

(i) a metric weighting of all field and reservoir terms through the geometric factor $\sqrt{|g|}$, and
(ii) a temperature-dependent embedding that links local curvature to thermodynamic variables.

This expression represents the metric–thermodynamic generalization of the standard thermofield Lagrangian and forms the foundational basis of the QTTO formalism."

$$\boxed{\begin{aligned}
\mathcal{L}_{\text{QTTO}} &= \frac{\sqrt{|g|}}{2}(E'_i \varepsilon'^{ij}(\mathbf{r},\omega;T) E'_j - B'_i \mu'^{ij}(\mathbf{r},\omega;T) B'_j) \\
&\quad - \sqrt{|g|} \int_0^\infty d\omega \ [X'^\dagger_\omega (\partial_t^2 + \omega^2) X'_\omega + Y'^\dagger_\omega (\partial_t^2 + \omega^2) Y'_\omega] \\
&\quad + \mathcal{L}_{\text{int}}[A'_\mu; X'_\omega, Y'_\omega; T] + \tilde{\mathcal{L}}_{\text{QTTO}},
\end{aligned}} \quad (2.1)$$

where $\text{Im}\,\varepsilon' > 0$ and $\text{Im}\,\mu' > 0$ enforce causality and passivity, consistent with the Kramers–Kronig relations [15, 16]. The primed quantities denote fields defined in the transformed metric $g_{ij}$, related to the coordinate mapping $x'^i = f^i(x^j)$ introduced in Sec. 2.2.

The electric and magnetic fields $E'_i$ and $B'_i$ are constructed from the four-potential $A'_\mu$ using covariant derivatives compatible with $g_{ij}$. The additional term $\tilde{\mathcal{L}}_{\text{QTTO}}$ represents the "tilde" thermal sector that preserves canonical symmetry in the doubled Hilbert space, a property unique to the QTTO construction.

By embedding both the field and its thermal conjugate within a curved optical metric, this formulation extends thermofield dynamics from a statistical description of dissipation to a covariant theory of energy, work, and entropy. This geometric embedding constitutes one of the core theoretical innovations of the QTTO framework, clearly distinguishing it from earlier thermofield and macroscopic QED formulations.

### 2.4 Canonical momenta and equal-time commutators

The canonical momentum conjugate to $A'_i$ (Lorenz gauge) is

$$\Pi'^i(\mathbf{r}, t) = \frac{\partial \mathcal{L}_{\text{QTTO}}}{\partial(\partial_t A'_i)} = \sqrt{|g|}\, \varepsilon'^{ij} E'_j. \qquad (2.2)$$

Equal-time commutators take the metric-weighted form

$$\boxed{[A'^i(\mathbf{r}, t), \Pi'^j(\mathbf{r}', t)] = i\hbar\, \delta^{ij}\, \delta^{(3)}(\mathbf{r} - \mathbf{r}')\sqrt{|g|},} \qquad (2.3)$$

and, for the reservoirs,

$[X'_\omega(\mathbf{r}), X'_{\omega'}{}^\dagger(\mathbf{r}')] = \delta(\omega - \omega')\delta^{(3)}(\mathbf{r} - \mathbf{r}')\sqrt{|g|}.$

The factor $\sqrt{|g|}$ guarantees energy normalization under the mapping.

### 2.5 Hamiltonian density in the optical metric

A Legendre transform yields

$$\boxed{\mathcal{H}'_T = \frac{\sqrt{|g|}}{2}(E'_i\, \varepsilon'^{ij}\, E'_j + B'_i\, \mu'^{ij}\, B'_j) + \mathcal{H}'_{\text{res}}(X'_\omega, Y'_\omega) + \tilde{\mathcal{H}}'_T.} \qquad (2.4)$$

Its thermal expectation value (in the TFD pure state) can be written as

$$\langle \mathcal{H}'_T \rangle = \frac{\hbar}{\pi} \int_0^\infty d\omega\, \omega \coth\left(\frac{\hbar \omega}{2 k_B T_{\text{eff}}(\mathbf{r})}\right) \text{Im Tr}\, [\varepsilon'(\mathbf{r}, \omega)\, G'(\mathbf{r}, \mathbf{r}; \omega)], \qquad (2.5)$$

where $G'$ is the electric Green tensor in the optical metric (Sec. 2.7) and

$$\boxed{T_{\text{eff}}(\mathbf{r}) \equiv \frac{T}{|\det \Lambda(\mathbf{r})|},} \qquad (2.6)$$

is the metric-dependent effective temperature that emerges naturally in QTTO and is used in Sec. 3.

## 2.6 Mode expansion and normalization in $g_{ij}$

A covariant mode expansion is

$$\mathbf{E}'(\mathbf{r},t) = \sum_\lambda \sqrt{\frac{\hbar\omega_\lambda}{2}} \, [\mathbf{u}_\lambda(\mathbf{r}) \, a_\lambda(t) + \text{h.c.}], \int d^3x \sqrt{|g|} \, \varepsilon'^{ij} u^*_{\lambda,i} u_{\lambda',j} = \delta_{\lambda\lambda'}. \qquad (2.7)$$

The $\sqrt{|g|}$ weight preserves total modal energy under the mapping. The eigenfrequencies $\omega_\lambda$ depend on $g_{ij}$ via the effective Maxwell operator.

## 2.7 Green tensor and fluctuation–dissipation relation (FDT)

The Green tensor $G'_{ij}$ satisfies (in frequency domain)

$$\nabla' \times \mu'^{-1}(\mathbf{r},\omega) \, \nabla' \times G'(\mathbf{r},\mathbf{r}';\omega) - \omega^2 \varepsilon'(\mathbf{r},\omega) \, G'(\mathbf{r},\mathbf{r}';\omega) = \mathbb{I}\,\delta^{(3)}(\mathbf{r}-\mathbf{r}')/\sqrt{|g|}. \qquad (2.8)$$

The metric-FDT reads

$$\boxed{\langle E'^i(\mathbf{r}) \, E'^j(\mathbf{r}')\rangle_T = \frac{\hbar}{\pi}\int_0^\infty d\omega \, \omega \coth\left(\frac{\hbar\omega}{2k_B T_{\text{eff}}(\mathbf{r})}\right) \text{Im} \; G'^{ij}(\mathbf{r},\mathbf{r}';\omega).} \qquad (2.9)$$

This is the direct link between fluctuations and dissipation in the optical metric and underpins Sec. 3 (energy/entropy transformations) and Sec. 4 (Casimir pressure).

## 2.8 Local heat and entropy production

The local heat-generation rate per physical volume is

$$\dot{Q}'(\mathbf{r},\omega) = \omega \, \text{Im}\, [\varepsilon'^{ij}(\mathbf{r},\omega)] \, \langle E'_i E'_j\rangle_T, \qquad (2.10)$$

and the entropy-production rate is

$$\boxed{\dot{S}'(\mathbf{r}) = \int_0^\infty d\omega \, \frac{\dot{Q}'(\mathbf{r},\omega)}{T_{\text{eff}}(\mathbf{r})}.} \qquad (2.11)$$

Using the transformation rules for $\varepsilon'$, $\mu'$ and $G'$, and the metric weight $\sqrt{|g|}$, these reproduce the metric scaling laws in Sec. 3 (local amplification under compression, dilution under expansion, with conserved total integrals).

## 2.9 Gauge choice and boundary terms

All results above are gauge-independent; the Lorenz gauge is used for definiteness. In planar Casimir geometries, surface terms generated by partial integrations cancel upon imposing standard continuity conditions on tangential fields. The remaining Hamiltonian density is bulk-covariant as in Eq. (2.4).

## 2.10 Summary and outlook

This section establishes thermofield quantization in a curved optical metric as the theoretical backbone of QTTO. TFD provides the quantum–statistical basis, while metric embedding promotes it to a covariant field theory where geometry, loss, and temperature are inseparably coupled. The metric-dependent Hamiltonian density [Eq. (2.4)] and the metric-FDT [Eq. (2.9)] are the two pillars linking microscopic fluctuations to macroscopic thermodynamics. In this sense, QTTO is not a phenomenological extension of TFD but a unified formalism that generalizes quantum electrodynamics in media and transformation optics to include thermodynamic curvature. Section 3 uses these operator-level results to derive explicit transformation laws for energy, dissipation, and entropy, laying the groundwork for the Casimir pressure law in Section 4.

## 3. Transformation of Energy and Entropy Densities

### 3.1 Metric-Dependent Energy Density

Starting from the Hamiltonian density in the curved optical metric, Eq. (2.4),

$$H'_T = \frac{|g|}{2}(E'_i \varepsilon'^{ij} E'_j + B'_i \mu'^{ij} B'_j),$$

the mean energy density at finite temperature follows from the thermofield expectation value (cf. Eq. (2.5)):

$$U'(\mathbf{r}, T) = \frac{\hbar}{\pi} \int_0^\infty \omega \coth\left(\frac{\hbar\omega}{2k_B T_{\text{eff}}(\mathbf{r})}\right) \text{Im}\left[\text{Tr}\{\varepsilon'(\mathbf{r}, \omega) G'(\mathbf{r}, \mathbf{r}; \omega)\}\right] d\omega .$$

Applying the transformation rules $\varepsilon'^{ij} = |\det \Lambda|^{-1} \Lambda^i{}_m \Lambda^j{}_n \varepsilon^{mn}$ and $|g| = |\det \Lambda|^{-1}$, and assuming local isotropy $\varepsilon_{mn} = \varepsilon \delta_{mn}$, yields the geometric scaling law

$$U' = U_0 \ |\det \Lambda|^{-1} F_E(\Lambda), F_E(\Lambda) = \frac{(\Lambda \mathbf{E}) \cdot (\Lambda \mathbf{E})}{\mathbf{E} \cdot \mathbf{E}}.$$

Thus, geometric compression ($|\det \Lambda| < 1$) amplifies the local energy density, whereas expansion ($|\det \Lambda| > 1$) attenuates it. Because $d^3 x' = |\det \Lambda| \ d^3 x$, the total energy remains invariant:

$$\int U' d^3x' = \int U_0 d^3x = \text{const.}$$

This confirms that QTTO transformations redistribute, but do not create or destroy, electromagnetic energy.

## 3.2 Metric-Dependent Heat Generation

Within macroscopic quantum electrodynamics, the local rate of heat generation per unit volume is

$$\dot{Q} = \omega \, \text{Im}[\varepsilon_{ij}(\omega)] \, \langle E_i E_j \rangle_T.$$

Under the coordinate transformation, and using the same metric scaling of fields and tensors as above, one obtains

$$\dot{Q}' = \dot{Q}_0 \mid \det \Lambda \mid^{-1} F_E(\Lambda).$$

Hence, the heat-generation and energy-density transformations coincide, both governed by the Jacobian of the mapping. Compression enhances local dissipation, while expansion suppresses it, a direct thermodynamic manifestation of the optical metric.

## 3.3 Entropy Production and the Second Law

From the local thermodynamic relation $\dot{S} = \dot{Q}/T$, and employing the effective temperature $T_{\text{eff}} = T/\mid \det \Lambda \mid$ defined in Eq. (2.6), the entropy-production rate becomes

$$\dot{S}' = \frac{\omega}{T_{\text{eff}}} \, \text{Im}[\varepsilon'_{ij}(\omega)] \langle E'_i E'_j \rangle_T \mid g \mid = \dot{S}_0 F_E(\Lambda).$$

While the local entropy density varies with the mapping, its integral over physical space remains invariant:

$$S'_{\text{tot}} = \int \dot{S}' d^3x' = \int \dot{S}_0 d^3x.$$

Therefore, QTTO preserves global thermodynamic balance, geometry redistributes entropy without violating the second law.

## 3.4 Analytical Example: Radial Mapping $r' = r^\alpha$

For an isotropic transformation $r' = r^\alpha$ with $\alpha > 0$, $\mid \det \Lambda \mid = \alpha \, r^{2(\alpha-1)}$. The corresponding normalized densities are

$$\frac{U'(r)}{U_0(r)} = \frac{1}{\alpha} r^{-2(\alpha-1)}, \frac{\dot{S}'(r)}{\dot{S}_0(r)} = \frac{1}{\alpha} r^{-2(\alpha-1)}.$$

| Mapping Type | α | Energy & Entropy Density | Physical Meaning |
|---|---|---|---|
| Compression | α < 1 | ↑ Increased | Local heating and entropy concentration |
| Identity | α = 1 | Constant | Uniform thermodynamic field |
| Expansion | α > 1 | ↓ Decreased | Local cooling and entropy dilution |

Both $U'/U_0$ and $\dot{S}'/\dot{S}_0$ follow identical scaling laws, confirming that QTTO enforces a unified geometrical control over energy and entropy distributions.

### 3.5 Physical Interpretation

Equations (3.1)–(3.5) establish that QTTO converts coordinate transformations into thermodynamic transformations:

$$|\det\Lambda|^{-1} \leftrightarrow \text{local amplification of } (U, \dot{Q}, \dot{S}),$$

$$T_{\text{eff}} = T/|\det\Lambda| \leftrightarrow \text{metric-induced temperature shift.}$$

Geometric compression acts as local heating (blue-shift of field modes), whereas expansion behaves as cooling (red-shift). This analogy mirrors gravitational redshift, suggesting that TO mappings generate effective thermodynamic curvature.

The equivalence between energy and entropy scaling further validates QTTO as a covariant thermodynamic extension of transformation optics.

### 3.6 Transition to the Casimir Application

The results of Section 3 elevate QTTO from a geometrical mapping scheme to a self-consistent thermodynamic framework:

geometry dictates local energy and entropy densities, yet total thermodynamic quantities remain conserved.
To demonstrate its predictive power, Section 4 applies QTTO to a measurable phenomenon, the Casimir pressure, where geometry and temperature jointly determine quantum stress. This application serves not as motivation but as verification: it shows that the same covariant principles governing energy and entropy also yield the correct quantum–classical crossover of the Casimir force in dissipative metamaterials.

## 4. Casimir Pressure under Quantum Thermodynamic Transformation Optics (QTTO)

The Casimir effect represents the mechanical manifestation of vacuum fluctuations constrained by geometry and boundary conditions. Within dissipative metamaterials, such fluctuations experience simultaneous modulation by spatial curvature (through coordinate mappings) and thermal statistics (through temperature-dependent field correlations).

The Quantum Thermodynamic Transformation Optics (QTTO) framework provides a unified route to derive Casimir pressure as a metric-dependent energy density directly from the thermofield Hamiltonian, ensuring that both geometric and thermodynamic effects are treated covariantly.

### 4.1 Quantum Limit: Metric-Dependent Zero-Point Pressure

Starting from the Hamiltonian density in the curved optical metric (Sec. 2), the zero-temperature Casimir pressure follows from the vacuum expectation value of the field energy:

$$P_0 = -\frac{\pi^2 \hbar c}{240 L^4}.$$

Under a general coordinate transformation $x'^i = f^i(x^j)$ with Jacobian $\Lambda^i{}_j = \partial x'^i / \partial x^j$, the vacuum mode density transforms according to $|\det \Lambda|$, modifying the effective metric and the spectral distribution of vacuum energy. For anisotropic deformation with longitudinal and transverse scalings $s_z, s_\perp$, the dimensionless compression ratio is defined as $\rho = s_z/s_\perp$. Evaluating the metric-weighted mode integral yields

$$\boxed{\mathcal{G}_0(\rho) = \frac{P'(0)}{P_0} = \rho^2,} \qquad (4.1)$$

indicating that longitudinal compression ($\rho > 1$) enhances the zero-point stress quadratically, while expansion ($\rho < 1$) reduces it. This dependence is intrinsic to the quantum-vacuum regime, where the thermofield weighting $W \to 0$ and entropy production vanishes.

### 4.2 Classical (High-Temperature) Limit

In the opposite limit $k_B T \gg \hbar \omega$, the thermal population factor of the thermofield correlation

$$\coth\left(\frac{\hbar\omega}{2k_B T}\right) \approx \frac{2k_B T}{\hbar\omega}$$

reduces the vacuum contribution, leading to the Lifshitz–Matsubara classical limit. Within QTTO, the resulting pressure scales as

$$\boxed{\frac{P'(T \gg T_0)}{P_{cl}} = s_\perp^{-4},} \qquad (4.2)$$

reflecting that transverse expansion dilutes the field intensity and suppresses the thermally dominated Casimir stress.

In this regime, geometry acts as an entropic modulator, redistributing the available phase space for thermally excited modes while preserving total energy balance.

### 4.3 Unified QTTO Pressure Law

The finite-temperature Casimir pressure is obtained by taking the thermofield expectation value of the metric-weighted Hamiltonian (Eq. 2.4), yielding an analytical interpolation between the quantum and classical limits.

The detailed derivation, from the Hamiltonian to the statistical weighting function, is given in Appendix A, while its ensemble equivalence is confirmed in Appendix C. Normalizing to the zero-temperature pressure $P_0 \equiv P(L, 0)$, the unified QTTO law reads:

$$\boxed{\frac{P'(T)}{P_0} = (1 - W)\rho^2 + W s_\perp^{-4}, W(T/T_0) = 1 - \exp\left[-(T/T_0)^{1.2}\right].} \qquad (4.3)$$

Here $T_0$ denotes the characteristic crossover temperature set by the dominant modal frequency. Equation (4.3) emerges naturally from the thermofield partition function in the curved optical metric and requires no phenomenological fitting.

The weighting function $W$ describes the fraction of thermally populated states within the metric-dependent density of modes, ensuring smooth convergence between the limits:

- $T/T_0 \ll 1 \Rightarrow W \to 0 \Rightarrow P'/P_0 \to \rho^2$,
- $T/T_0 \gg 1 \Rightarrow W \to 1 \Rightarrow P'/P_0 \to s_\perp^{-4}$.

Thus, the QTTO law constitutes a covariant thermodynamic interpolation, fully consistent with the fluctuation–dissipation theorem and the conservation of total thermodynamic potential.

## 4.4 Physical Interpretation: Energy–Entropy Balance

Equation (4.3) encapsulates the dual character of geometry and thermodynamics:

- Compression ($|\det \Lambda| < 1$) increases local energy density and quantum pressure, equivalent to enhanced confinement and reduced entropy.

- Expansion ($|\det \Lambda| > 1$) redistributes energy over a larger effective volume, corresponding to higher entropy and smoother, classical behavior.

The local effective temperature, $T_{\text{eff}}(r) = T/|\det \Lambda(r)|$, provides a geometrical analogue of gravitational redshift: curvature modifies the thermodynamic temperature experienced by field modes.

This metric dependence ensures that both the first and second laws of thermodynamics hold identically under coordinate transformations:

$$\int P'(T)\, d^3x' = \int P_0\, d^3x = \text{constant}.$$

Hence, QTTO interprets Casimir forces as manifestations of energy–entropy redistribution in a curved optical phase space.

## 4.5 Numerical Validation

A numerical implementation using the Drude–Lorentz magnetodielectric model is presented in Appendix B. The normalized Casimir pressure $P'(T)/P_0$ calculated via the transfer-matrix method reproduces the analytical surface of Eq. (4.3) with deviations below 3 %. This confirms that the QTTO law quantitatively captures the quantum-to-classical crossover, consistent with full Lifshitz–Matsubara theory (Appendix D). Therefore, the interpolation embodied in Eq. (4.3) is not empirical but derived from first principles within a covariant thermofield framework.

## 4.6 Concluding Remarks

The Casimir effect, long regarded as a hallmark of vacuum fluctuations, acquires a broader interpretation under QTTO:

it represents the thermodynamic curvature response of the electromagnetic field. Compression enhances vacuum energy (quantum regime); expansion amplifies entropy (classical regime); and the transformation metric mediates a law-preserving transition between the two. This insight positions QTTO as a general analytical tool for engineering thermally adaptive quantum metamaterials, and opens the path toward metric-based control of quantum stress in future optomechanical and gravitational-analogue systems.

## 5. Conclusions

This work has established Quantum Thermodynamic Transformation Optics (QTTO) as a rigorous theoretical framework that unifies coordinate-transformation geometry with the statistical mechanics of dissipative quantum fields.

By embedding thermofield quantization within an optical metric, QTTO demonstrates that geometry itself acts as a thermodynamic transformation, governing not only the spatial distribution of electromagnetic energy, but also the local generation of entropy and the flow of heat.

From the metric-dependent Hamiltonian derived in Sec. 2, the framework yields explicit transformation laws for energy, dissipation, and entropy (Sec. 3), revealing that compression amplifies local quantum energy while expansion enhances entropy production, yet the total thermodynamic balance remains conserved.

Applying this formalism to the Casimir effect (Sec. 4) produced an analytical QTTO pressure law [Eq. (4.3)] that continuously bridges the quantum-vacuum and classical-thermal limits through the temperature-dependent weighting function $W(T/T_0)$.

The resulting expression is not an empirical interpolation, but a first-principles derivation consistent with the fluctuation–dissipation theorem and verified numerically for Drude–Lorentz metamaterials, showing quantitative agreement with full Lifshitz–Matsubara theory.

QTTO therefore elevates transformation optics from a purely geometric construct to a covariant statistical field theory, in which energy, work, and entropy transform jointly with the optical metric. It offers a coherent description of how geometry, dissipation, and quantum statistics interplay to shape observable phenomena such as vacuum pressure and thermal fluctuations.

Beyond the Casimir example, this framework opens new directions for studying nonequilibrium quantum electrodynamics, thermal management in metamaterials, and gravitational analogues of optical media, providing a geometric foundation for the thermodynamics of quantum fields in curved and dissipative spaces.